\documentclass[sn-mathphys,Numbered]{sn-jnl}% Math and Physical Sciences Reference Style
\usepackage{graphicx}%
\usepackage{multirow}%
\usepackage{amsmath,amssymb,amsfonts}%
\usepackage{amsthm}%
\usepackage{mathrsfs}%
\usepackage[title]{appendix}%
\usepackage{xcolor}%
\usepackage{textcomp}%
\usepackage{manyfoot}%
\usepackage{booktabs}%
\usepackage{algorithm}%
\usepackage{algorithmicx}%
\usepackage{algpseudocode}%
\usepackage{listings}%
\usepackage{indentfirst}%
\usepackage{siunitx}%
\usepackage{lineno}
\theoremstyle{thmstyleone}%
\theoremstyle{thmstyletwo}%

\theoremstyle{thmstylethree}%

\newcommand{\K}{\mbox{$^{40}$K}}

\newcommand{\Lu}{\mbox{$^{176}$Lu}}
\newcommand{\V}{\mbox{$^{50}$V}}
\newcommand{\La}{\mbox{$^{138}$La}}
\newcommand{\Te}{\mbox{$^{123}$Te}}
\newcommand{\Ta}{\mbox{$^{180}$Ta}}
\newcommand{\Yb}{\mbox{$^{176}$Yb}}
\newcommand{\Hf}{\mbox{$^{176}$Hf}}

\newcommand{\TvBB}{$2\nu\beta\beta$}
\newcommand{\ZvBB}{$0\nu\beta\beta$}

\begin{document}

\title[Article Title]{LUCE: A milli-Kelvin calorimeter experiment to study the electron capture of \Lu}


\author[1]{\fnm{Shihong} \sur{Fu}}

\author[2,3]{\fnm{Giovanni} \sur{Benato}}

\author[3]{\fnm{Carlo} \sur{Bucci}}

\author[3]{\fnm{Paolo} \sur{Gorla}}

\author[3]{\fnm{Pedro V.} \sur{Guillaumon}}\equalcont{Corresponding author: pedro.guillaumon@lngs.infn.it }%\email{pedro.guillaumon@lngs.infn.it}

\author[4]{\fnm{Jiang} \sur{Li}}

\author[5,8]{\fnm{Serge} \sur{Nagorny}}

\author[6,7]{\fnm{Francesco} \sur{Nozzoli}}

\author[2,3]{\fnm{Lorenzo} \sur{Pagnanini}}

\author[3]{\fnm{Andrei} \sur{Puiu}}

\author[2,3]{\fnm{Matthew} \sur{Stukel}}

\equalcont{Corresponding authors: pedro.guillaumon@lngs.infn.it, matthew.stukel@gssi.it }

\affil[1]{Fudan University, China}

\affil[2]{Gran Sasso Science Institute, Italy}

\affil[3]{Laboratori Nazionali del Gran Sasso, INFN, Italy}

\affil[4]{Shanghai Institute of Ceramics, Chinese Academy of Science, China}

\affil[5]{Department of Physics, Engineering Physics and Astronomy, Queen's University, 64 Bader Lane, K7L 3N6 Kingston, ON, Canada
}

\affil[6]{INFN-TIFPA, Italy}

\affil[7]{Department of Physics, Trento University, Italy}

\affil[8]{Arthur B. McDonald Canadian Astroparticle Physics Research Institute, 64 Bader Lane, K7L 3N6, Kingston, ON, Canada}
%\affil[3]{\orgdiv{Department}, \orgname{Organization}, \orgaddress{\street{Street}, \city{City}, \postcode{610101}, \state{State}, \country{Country}}}


\abstract{ The LUCE (\textbf{LU}tetium  s\textbf{C}intillation \textbf{E}xperiment) project will search for the \Lu\ electron capture based on a milli-Kelvin calorimetric approach. This decay is of special interest in the field of nuclear structure, with implications for the s-process and for a better comprehension of the nuclear matrix elements of neutrinoless double beta decay ($\mathrm{0\nu\beta\beta}$) and two-neutrino double beta decay ($\mathrm{2\nu\beta\beta}$).
Possible impacts also include the development of a new class of coherent elastic neutrino-nucleus scattering ($\mathrm{CE\nu NS}$) and spin-dependent (independent) dark matter detectors. We report on the current status and design of a novel detector cryogenic-module for the measurement of the electron capture and detail a future measurement plan. }




\maketitle

\section{Introduction}\label{sec:Introduction}

The study of rare decays is of fundamental importance in the field of nuclear and particle astrophysics. These decays allow for a better understanding of the nuclear structure involved in these processes by extracting the electroweak coupling quenching to observe if any violation occurred when comparing the predictions of the Standard Model~\cite{electroweak,electroweak2}. There are two main implications in these studies: a better prediction of \TvBB\ and \ZvBB\ half-lives and of the nucleosynthesis of s-process \cite{lu176-1,lu176-2,lu176-3,lu176-4}.

There are only six, known, long-lived (over a billion years) nuclides that undergo an electron capture (EC) decay, namely \K\ (1.29$\times$10$^{12}$~yrs~\cite{stukel2022rare,hariasz2022first}), \V\ (2.2$\times$10$^{17}$~yrs~\cite{v50,laubenstein2019new}), \La\ (1.6$\times$10$^{11}$~yrs~\cite{firestone}), \Te\ ($>$10$^{16}$-10$^{19}$~\cite{te123_1,te123_2}), \Ta\ ($>$2.0$\times$10$^{17}$~yrs~\cite{ta180,cerroni2023deep}), and \Lu\ ($>$10$^{13}$-10$^{14}$~yrs~\cite{francesco}), three of them have never been detected. The goal of the LUCE experiment (\textbf{LU}tetium  s\textbf{C}intillation \textbf{E}xperiment) is to measure the partial half-life of \Lu\ with a milli-Kelvin scintillating calorimeter in an ultra-low-background facility. Using low-temperature calorimeters is a new approach to this measurement and will lead to high sensitivity~\cite{NORMAN2004767, francesco}.

\begin{figure}
    \centering	\includegraphics[width=0.5\linewidth]{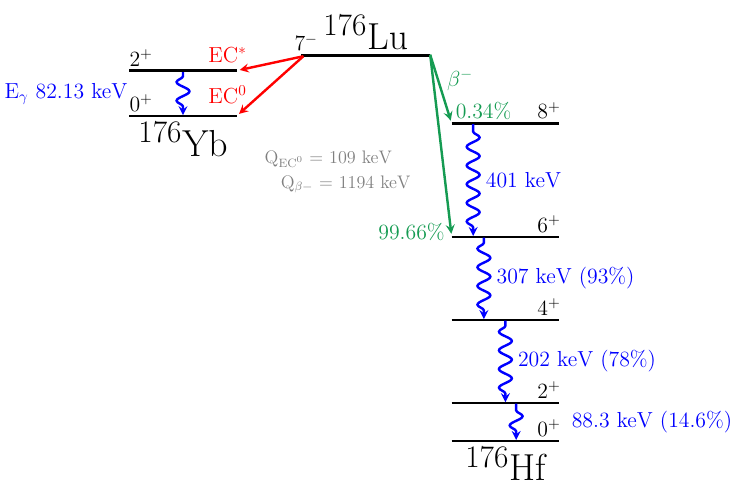} 
	\caption{Decay scheme of $^{176}$Lu.~\cite{francesco}}
    \label{Fig:Lu176_Decay_Scheme}
    \end{figure}

The dominant \Lu\ ground state (g.s.) decay is a $\beta^-$-decay ($J^\pi$(\Lu) = 7$^-$ $\rightarrow$ $J^\pi$(\Hf) = 0$^+$) with a half-life of $(3.5-3.8) \times 10^{10}$ years and is  used in dating methods of meteorites and rocks \cite{NORMAN2004767}. The precise value for this half-life has been subject to discussion, and one of the possible explanations is the existence of an EC decay to \Yb. The EC decay is energetically allowed, having a Q-value of $109$~keV to the \Yb\ g.s. ($J^\pi$(\Lu) = 7$^-$ $\rightarrow$ $J^\pi$(\Yb) = 0$^+$) and $\sim 27$~keV to the first excited state ($J^\pi$(\Lu) = 7$^-$ $\rightarrow$ $J^\pi$(\Yb) = 2$^+$). These decays, however, would be $\mathrm{7^{th}}$ and $\mathrm{5^{th}}$ forbidden decays and have never been detected. Fig.\ref{Fig:Lu176_Decay_Scheme} shows the current decay scheme for \Lu.

\section{The LUCE detector concept}\label{sec:Detector_Concept}
\begin{figure}
    \centering	\includegraphics[width=0.5\linewidth]{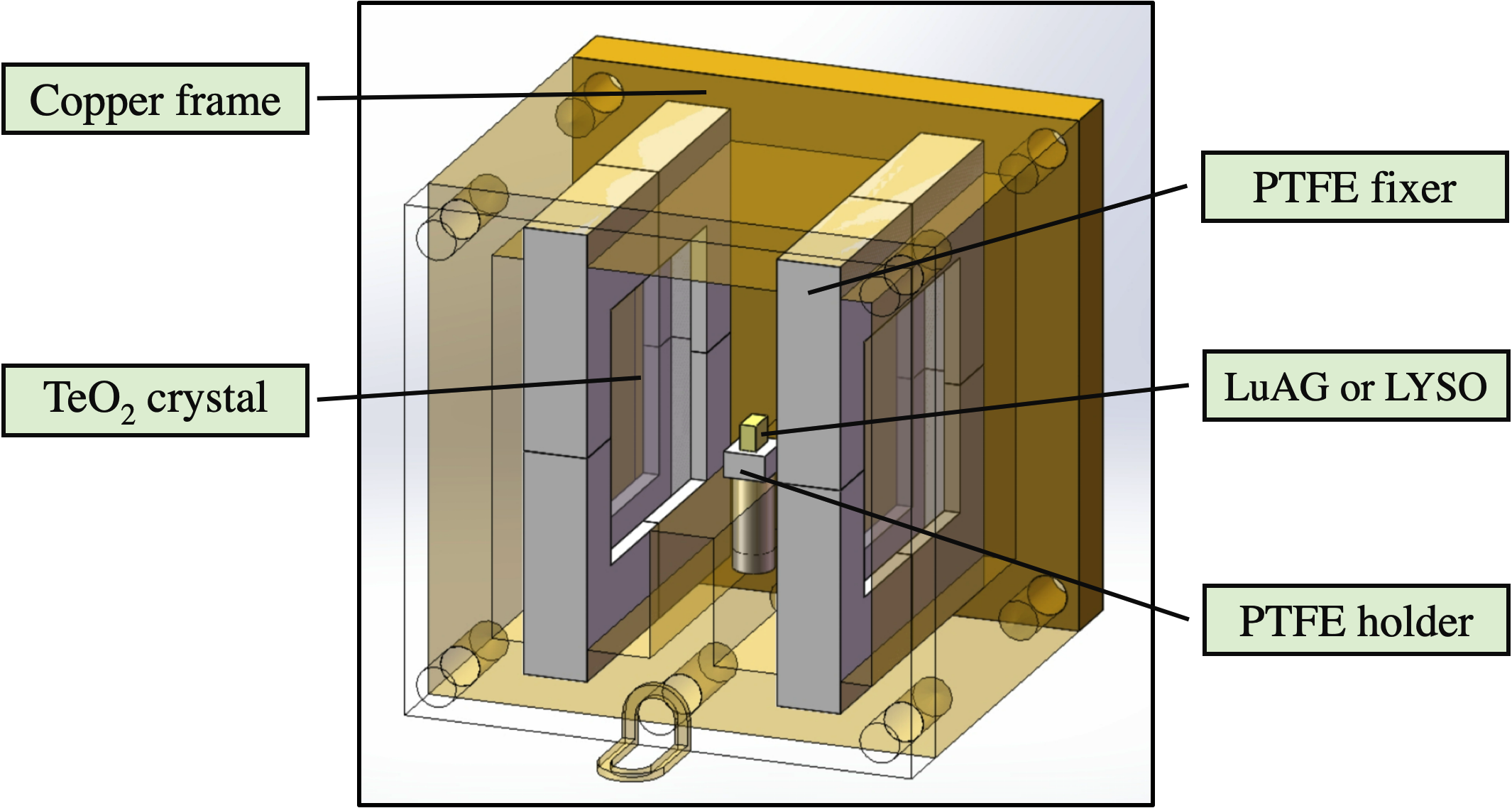}
	\caption{ Proposed detector module for the LUCE experiment.}
    \label{Fig:Lu176_holder}
    \end{figure}

LUCE is a proposed experiment for the detection of the \Lu\ EC-decay based on an innovative cryogenic calorimetric technique operated at $\sim 15$~mK. It relies on a modular design where a LYSO [$\mathrm{(Lu,Y)_2SiO_5}$:Ce]  or LuAG [$\mathrm{Lu_3Al_5O_{12}}$:Ce] crystal scintillator containing the \Lu\ source is placed in the middle of a copper box and is surrounded by two \mbox{TeO${}_{2}$} crystals, shown in Fig.~\ref{Fig:Lu176_holder}. When \Lu\ decays, its  products can be fully or partially absorbed in the crystal, producing a variation of temperature. This change in temperature can be measured by an attached NTD-Ge sensor~\cite{haller1984ntd}. The raw signal will be saved separately for digital post-processing and anti-coincidence analysis.

The main source of background is the \Lu\ $\beta$-decay. More than 99\% of \Lu\ decays to the $6^+$ state which has an energy of 596.82 keV. This means that the rest of the Q-value (593.88 keV) goes to a continuum distribution of the electron. Due the small size of the crystal and depending on the energy of the electron, the particle can deposit just a fraction of the energy inside the crystal, giving rise to a distorted $\beta$-shape distribution. This background can be reduced by running the TeO$_2$ crystals in coincidence with the \Lu\ source. The higher energy gammas produced during the $\beta$-decay will escape from the \Lu\ source crystal unlike the products of the EC. Preliminary simulations, for a 4$\times$4$\times$20~cm$^3$ crystal, show that $\sim$50$\%$ of the beta events can be removed by the two (50 $\times$ 50 $\times 5~\mathrm{mm}^3$) $\mathrm{TeO}_2$ crystals placed 15~mm from the \Lu\ source crystal. While greater than 90$\%$ of the electron capture events will remain inside. Further simulations will be performed to optimize the experimental geometry.

In order to avoid pile-ups due to the long decay time in the NTD, the lutetium crystals will need to have an activity of less than $<$1~Bq, which means crystals size of $\sim$1~mm$^3$ and smaller sensors matching the thermal impedance. With a crystal of this size, it is expected to have sensitivity to a decay of $10^{14}$ years, after 1 month of measurement, see Sec.~\ref{subsec:Sensitivity} for further details. LuAG and LYSO crystals will be tested as cryogenic calorimeters. 

The $\mathrm{5^{th}}$ forbidden EC decay to the 2$^+$ excited state produces a 82.1 keV de-excitation photon. Due to the rising time of the NTD, the de-excitation photon will be visible with the Auger or x-rays produced from the atomic re-arrangement of the \Yb\ (K transition is energetically prohibited), resulting in a peak search at 82.1 + 10.481 (L1 binding energy) = 92.6~keV. The 7$^{th}$ forbidden EC g.s. decay allows for both the K and L captures of \Yb~\cite{NORMAN2004767} (they are the most probable) resulting in peaks at 61.3 keV (log ft = 0.532) and 10.5 (log ft = 0.347) keV respectively \cite{lu-teor}.

\section{Experimental Program}\label{sec:Exp_Program}

The goal of LUCE phase-I is to have one module fully operational and reach or exceed the current literature half-life sensitivity. Using the knowledge gained during phase-I, LUCE phase-II will design an experiment that will aim to discover or provide a world leading sensitivity for the electron capture. In phase-I the detector response (both phonon propagation and light yield) at low-temperature must be studied. LuAG:Ce and LYSO:Ce crystals will be studied in order to choose the one with higher resolution in the region-of-interest (ROI), in the heat channel. Although light yield measurements are not critical to measure the EC of \Lu, this will also allow to characterize this crystal for other applications in neutrino and dark matter physics. A detailed simulation program is being developed concurrently to support experimental measurements. The following sections outline the current status of the experiment and details future operations.

\subsection{Light Yield Measurement}\label{subsec:Light_Yield}

\begin{figure}
    \centering	\includegraphics[width=0.5\linewidth]{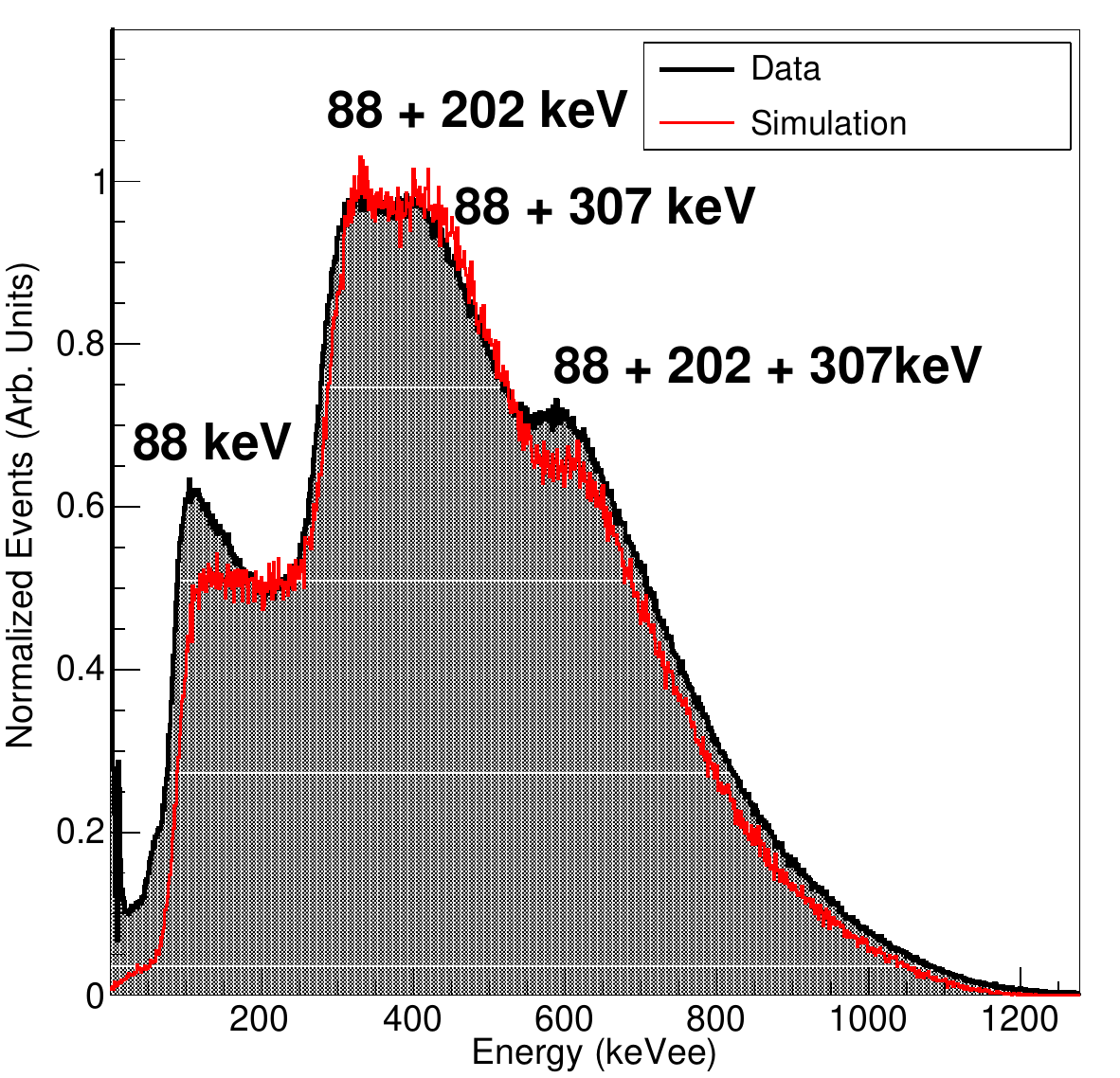} 
	\caption{Intrinsic energy spectrum of a LYSO crystal along with the simulated spectrum. The different $\beta$-$\gamma$-ray/internal conversion combinations allowed during the decay of \Lu\ are also shown. }
    \label{Fig:LYSO_Spectrum.pdf}
    \end{figure}

Light yield measurements as a function of temperature, for both crystals, will be performed in order to have a full understanding of low-temperature detector response. A room-temperature spectral measurement was performed with a 4$\times$4$\times$20~mm$^{3}$ LYSO crystal and photomultiplier tube (PMT) to see the intrinsic radiation and to perform a first comparison with the simulations. The energy spectrum with simulation comparison can be seen in Fig.~\ref{Fig:LYSO_Spectrum.pdf}. The data was collected over 16 hours in a surface facility with a hardware trigger where the data points represent the pulse height. In general, the simulated data follows the shape of experimental data. There is some discrepancy at low and high energy but this is most likely explained by the light collection efficiency, Birks quenching and pile-up, which is not yet included in the simulation. The measured spectrum shows the shifting of the $\beta^-$ spectrum by the different combinatorics of the gamma rays (i.e. 88.3~keV, 290.3~keV (88.3 + 202~keV), and 395.3~keV (88.3 + 307~keV)). The spectral shape and subsequent intensities of the $\beta^-$ peaks will ultimately be dependent on the final crystal size selected.   

We have developed a system to measure the scintillation characteristics of crystals, such as LYSO and LuAG, over a wide temperature range (\qtyrange[range-phrase = --]{10}{300}{\K}) using a Xenon (Xe) arc lamp as the excitation source. This method has been shown to determine the relationship between light yield and temperature in previous studies on similar crystals~\cite{KIM2021109396}. The results are expected to reveal a detailed temperature dependence of light yield, which is crucial for optimizing the detector performance at cryogenic temperatures for other applications.

\subsection{Calorimetric Measurements}\label{subsec:Calorimeter}

A small sample with an attached NTD-Ge sensor, of LuAG (grown by the Shanghai Institute Of Ceramics Chinese Academy Of Sciences) and of LYSO (grown by Saint Gobain), will be tested in the underground facility of Laboratori Nazionali del Gran Sasso. The primary goal is to show these crystals can work as low-temperature calorimeters and determine their low-energy threshold and resolution. One of the critical aspects of the LUCE project is the ability of larger crystals to handle the high counting rate of intrinsic radioactivity present in the sample and avoid pile-up. In that way, the evaluation of rise and decay times of pulses is crucial. 

LYSO crystals have a percentage of yttrium that varies from 5\% to 70\% \cite{Knoll:1300754}. Depending on the amount of yttrium, the arrangement of the crystal lattice will change affecting the phonon propagation, and, by consequence, the resolution of the crystal. Although not a perfect comparison (difference in size and quality), garnet crystals similar to LuAG have already been tested as cryogenic calorimeters, showing energy threshold lower than 8 keV with good resolution pulses \cite{tm_garnet}. There is a lack of information about the heat capacity of LuAG and LYSO crystals, but since the specific heat at room temperature of those crystals, are comparable to other low-temperature detectors, it is expected they will behave well as cryogenic detectors \cite{Kuwano2004,cong2009structural,Musikhin2015}. Also, no high paramagnetism have been reported in LuAG and LYSO crystals. Studies on the thermal properties of those crystals will need to be performed.

\subsection{Projected Sensitivity}\label{subsec:Sensitivity}

\begin{figure}
    \centering	\includegraphics[width=1.0\linewidth]{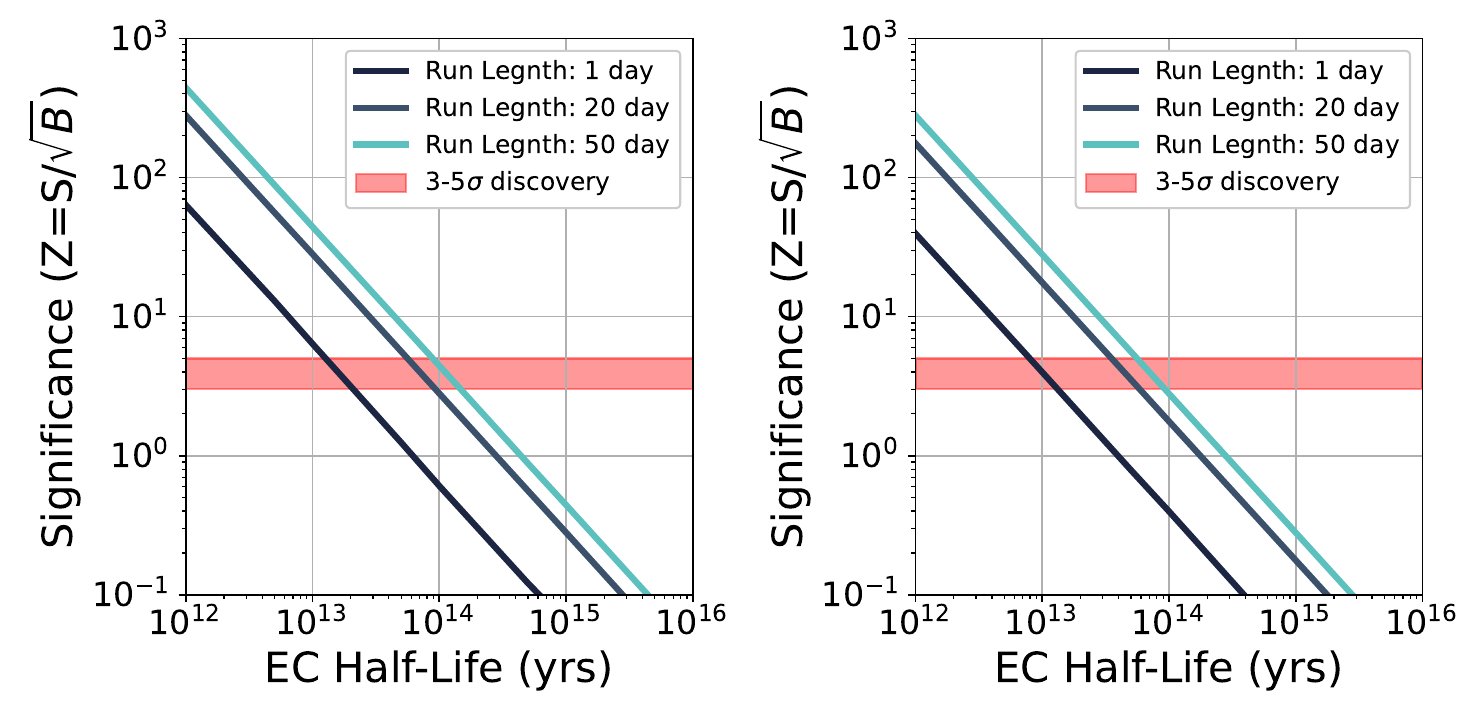} 
	\caption{Significance of the LUCE experiment. The significance is defined as the signal (S) over the square root of the background (B). Current experimental lower limits are at 10$^{13}$-10$^{14}$ years~\cite{francesco}. (Left) Ground state electron capture significance (61.3~keV). (Right) Excited state electron capture significance (92.6~keV). }
    \label{Fig:LUCE_Sensitvity_Duo.pdf}
    \end{figure}

Based on the detector setup proposed in Sec.~\ref{sec:Detector_Concept} the significance of a single module was studied. This study simulated a 1$\times$1$\times$2.5~mm$^{3}$ LuAG crystal with the resolution from~\cite{lu-last}. The signal was generated as a gaussian at ground (61.3~keV) and excited state (92.6~keV). The significance was evaluated at each of these energies. The number of signal events depended on the selected half-life for the process. The only background considered was the events from the $\beta^-$ decay that were not found in coincidence with the adjacent TeO$_2$ crystals. This would be the dominant background in an underground run. The significance was evaluated as the signal counts over the square-root of the background in a selected energy region, $\pm$10~keV around the above peak means. The significance as a function of half-life and run time can be seen in Fig.~\ref{Fig:LUCE_Sensitvity_Duo.pdf}. Here it is shown that this small crystal can reach the current literature sensitivities with two months of run time. Improvement to the detectable half-life or significance are easily achievable by increasing the size of the crystal or the number of detector modules in the experiment. Additionally, the resolution of the calorimeter will effect the sensitivity. It is expected that the initial cryogenic tests will show that the actual resolution will be smaller than the one used in this study. It has been reported a FWHM of 7 keV at 3 MeV with NTD-Ge macrocalorimeters \cite{nature-2022} and of 3.1 eV at 6keV with NTD-Ge microcalorimeters \cite{silver2005ntd}, which is  better than other, typical, experimental setup.

\section{Conclusion and Perspectives}

The LUCE project is the first experiment to search for the EC-decay of \Lu\ by using a lutetium based cryogenic calorimeter. The modular and flexible detector concept is designed to handle the high intrinsic activity and can be repurposed for other experiments. LUCE phase-I will be the proof-of-concept for the design and will have the sensitivity to reach or exceed current literature values of the EC-decay half-life ($\sim 10^{14}$ years) in a two month data taking campaign. Phase-II is expected to have multiple, optimized detector modules to pursue discovery of the branching ratio or achieve world leading limits. The LUCE experiment is a unique application of underground particle physics techniques applied to nuclear physics problems and could open a new frontier of precision nuclear measurements.

\bibliography{sn-bibliography}% common bib file

\end{document}